\documentclass[pra,twocolumn,showpacs,groupedaddress,superscriptaddress,aps,10pt]{revtex4-1}
\usepackage{bm,graphicx,amsmath}
\usepackage{placeins}
\usepackage{amssymb}
\usepackage{amsmath}
\usepackage{epsfig}
\usepackage{amssymb}
\usepackage{color}
\usepackage[colorlinks,linkcolor=blue,citecolor=blue]{hyperref}
\usepackage{subfigure}

\begin{document}

\title{Super-Gaussian conical refraction beam}
\date{\today}

\author{A. Turpin}
\affiliation{Departament de F\'isica, Universitat Aut\`onoma de Barcelona, Bellaterra, E-08193, Spain}
\author{Yu. V. Loiko}
\affiliation{Departament de F\'isica, Universitat Aut\`onoma de Barcelona, Bellaterra, E-08193, Spain}
\affiliation{Aston Institute of Photonic Technologies, School of Engineering \& Applied Science Aston University, Birmingham, B4 7ET, UK}
\author{T. K. Kalkandjiev}
\affiliation{Departament de F\'isica, Universitat Aut\`onoma de Barcelona, Bellaterra, E-08193, Spain}
\affiliation{Conerefringent Optics SL, Avda. Cubelles 28, Vilanova i la Geltr\'u, E-08800, Spain}
\author{H. Tomizawa}
\affiliation{JASRI/SPring-8, 1-1-1, Kouto, Sayo, Hyogo, 679-5198, Japan}
\author{J. Mompart}
\affiliation{Departament de F\'isica, Universitat Aut\`onoma de Barcelona, Bellaterra, E-08193, Spain}

\begin{abstract} 
We demonstrate the transformation of Gaussian input beams into super-Gaussian beams with quasi flat-top transverse profile by means of the conical refraction phenomenon by adjusting the ratio between the ring radius and the waist radius of the input beam to 0.445. We discuss the beam propagation of the Super-Gaussian beam and show that it has a confocal parameter three times larger than the one that would be obtained from a Gaussian beam. The experiments performed with a ${\rm KGd(WO_4)_2}$ biaxial crystal are in good agreement with the theoretical predictions.  \\
\textbf{ocis}:(140.3300) Laser beam shaping, (160.1190) Anisotropic optical materials; (260.1180) Crystal optics; (260.1140) Birefringence; (350.3390) Laser materials processing. 
\end{abstract}

\date{\today}
\maketitle
Ideally, a flat-top beam \cite{bagini1996,gur1998,tovar2001,cai2006,dan2008,cai2008,liu2009,forbes2009,ma2010,jahn2010,zhan2011,forbes2012,forbes2013,zhan2009,jr2014} is a light beam possessing an intensity transverse profile mostly flat in the central part and sharply decaying at its edges, at variance with the Gaussian profile of the fundamental TEM$_{00}$ mode. Flat-top beams are useful in a wide variety of laser applications where one needs a uniform intensity over a fixed area, such as in optical processing \cite{zhan2009,roach1979}, laser-driven acceleration of particles \cite{wang2007, barker2012}, optical trapping \cite{china2009} or gravitational-waves detectors \cite{ju2010}.  
Nevertheless, the generation of flat-top beams is a non trivial task and usually diffractive optical elements are required, which suffers from several drawbacks such as losses due to inefficient mode projection or diffraction, their extreme precise control or their limited spectral range \cite{zhan2009}. Beams whose transverse intensity profile possess sharp edges and extremely flat cross-section are ideal realizations of flat-top beams. In experimental situations flat-top beams are well approximated by \textit{super-Gaussian} beams \cite{forbes2003,preeza2009,kim2012,book2000}. A super-Gaussian beam also possesses a flat intensity profile but it decays smoothly at its edges, similarly to a Gaussian beam \cite{book2000}. The aim of this work is to show that a beam of super-Gaussian profile can be generated by transforming an input Gaussian beam with a biaxial crystal throughout the conical refraction (CR) phenomenon \cite{1978_Belskii_OS_44_436,1999_Belsky_OC_167_1,2004_Berry_JOA_6_289,2007_Berry_PO_55_13,2008_Kalkandjiev_SPIE_6994,ebs2013,2013_Sokolovskii_OE_21_11125}. 

\begin{figure}[]
\centering
\includegraphics[width=\columnwidth]{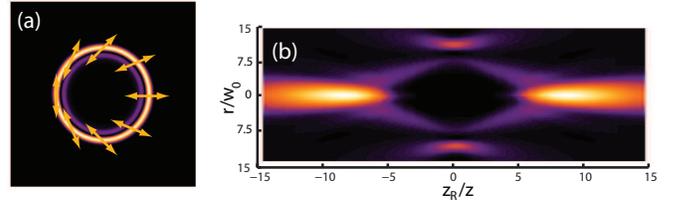}
\caption{Fundamental features of the CR phenomenon under conditions of $\rho_0 \equiv R_0 / w_0 \gg 1$. (a) Bright and dark Poggendorff rings and field polarization distribution (yellow double arrows). (b) Intensity distribution in the $z$-$r$ plane showing the free space evolution of the CR beam.}
\label{fig1}
\end{figure}

CR in biaxial crystals (BCs) is an old phenomenon predicted theoretically by Hamilton in 1832 and demonstrated experimentally by Lloyd few months later \cite{2007_Berry_PO_55_13}. CR is usually described in terms of the transformation of an input Gaussian beam into a bright ring when the first one propagates along one of the optic axes of a BC. The radius of this light ring, $R_0$, is the product of the length of the BC, $l$, and its birefringence or conicity parameter, $\alpha$, i.e. $R_0 = l \alpha$. One of the signatures of the CR phenomenon is the polarization distribution along the ring, which differs from the well-known radial and azimuthal modes, so that every pair of diametrically opposite points possess orthogonal polarizations. This polarization distribution only depends on the orientation of the plane of optic axes of the BC \cite{2008_Kalkandjiev_SPIE_6994}. Another specific characteristic of the light ring of CR is that, for $\rho_0 \equiv R_0 / w_0 \gg 1$ (where $w_0$ is the waist radius of the focused input beam), it splits at the focal plane into a pair of bright concentric rings. As the imaging plane is symmetrically moved from the focal plane, the transverse pattern becomes more complex, including secondary rings. At a distance $Z_{Raman} = \pm \sqrt{4/3} z_R \rho_0$ most of the light gets concentrated in two (Raman) spot, where $z_R = \pi w_0^2 / \lambda$ is the Rayleigh range of the focused input beam \cite{vault}. All these features are sketched in Fig.~\ref{fig1}. 

\begin{figure}[]
\centering
\includegraphics[width=\columnwidth]{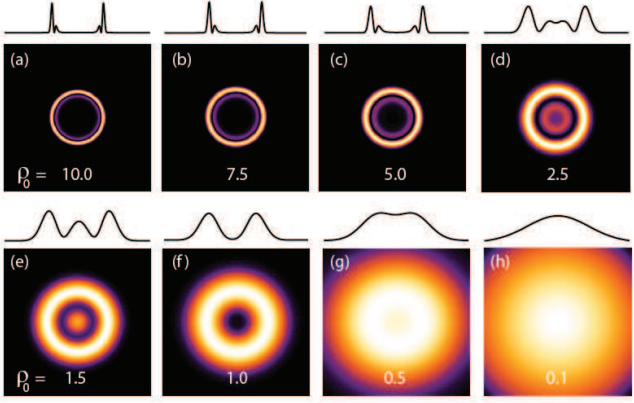}
\caption{Transverse intensity distributions at the focal plane $I_{\rm{CP}} \left( \rho, Z=0 \right)$ for an input Gaussian beam of circular polarization as given by Eqs.~(\ref{eqI})-(\ref{Eqs_Bs_general}) for different values of the control parameter $\rho_{0}$.}
\label{fig2}
\end{figure}

The diffraction theory of CR was a result of the investigation of many authors that was finally recapitulated by Belsky, Khapalyuk and Berry \cite{1978_Belskii_OS_44_436, 1999_Belsky_OC_167_1, 2004_Berry_JOA_6_289}. From this theory it follows that for a circularly polarized input beam, the light's intensity distribution behind the BC is given by:
\begin{equation}
I(\rho, Z)= \left| B_C \right|^2 + \left| B_S \right|^2~, 
\label{eqI}
\end{equation}
where 
\begin{eqnarray}
B_{C}\left( \rho,Z \right) = \int_{0}^{\infty }  \eta a \left( \eta \right)
e^{-i \frac{Z}{4} \eta ^{2} } \cos \left( \eta \rho_{0}\right) J_{0}\left( \eta \rho \right) d \eta, 
\label{Eqs_Bc_general} \\
B_{S}\left( \rho,Z \right) = \int_{0}^{\infty }  \eta a \left( \eta \right)
e^{-i \frac{Z}{4} \eta ^{2} } \sin \left( \eta \rho_{0}\right) J_{1}\left( \eta \rho \right) d \eta, 
\label{Eqs_Bs_general} \\
a \left( \eta \right) = \int_{0}^{\infty} \rho E_{\rm{in}} \left( \rho \right) J_{0} \left( \eta \rho \right) d \rho. \label{Eqs_aP_uniform}
\end{eqnarray}
In these equations $\rho = r / w_0$ and $Z = z / z_R$, being $r$ and $z$ the radial and longitudinal components in cylindrical coordinates. $a \left( \eta \right)$ is the radial part of the 2D transverse Fourier transform of the input beam $\mathbf{E}_{\rm{in}}=E_{\rm{in}} \left( \rho \right) \mathbf{e}_{\rm{in}}$, $\eta = \left| \mathbf{k}_{\perp} \right| w_{0}$ is the modulus of the transverse wave-vector components projected onto the entrance surface of the crystal, and $J_{q}$ denotes the Bessel function of order $q$. For an input Gaussian beam of fundamental transverse profile, $a \left( \eta \right) = \exp \left( -\eta ^{2} / 4 \right)$ \cite{2004_Berry_JOA_6_289}. Since the role of the factor $\rho_0$ is crucial in the resulting pattern of CR, we use it as our control parameter. For the case $\rho_0 \gg 1$ where the two well resolved CR concentric rings appear, CR has lead to applications in the fields of optical trapping \cite{vault}, lasers \cite{loiko2014}, free space optical communications \cite{fsoc}, polarimetry \cite{polarimetry} and mode conversion \cite{peet2010jo}, by taking profit of the well defined CR rings and their polarization. However, when one moves to a configuration where $\rho_0 \approx 1$, the transverse pattern of CR becomes substantially different, see Fig.~\ref{fig2}. This region has been explored recently, showing that CR can be used to improve the directivity of laser beams \cite{peet2010} and also to generate a three dimensional dark focus \cite{hole}. 

\begin{figure}[]
\centering
\includegraphics[width=\columnwidth]{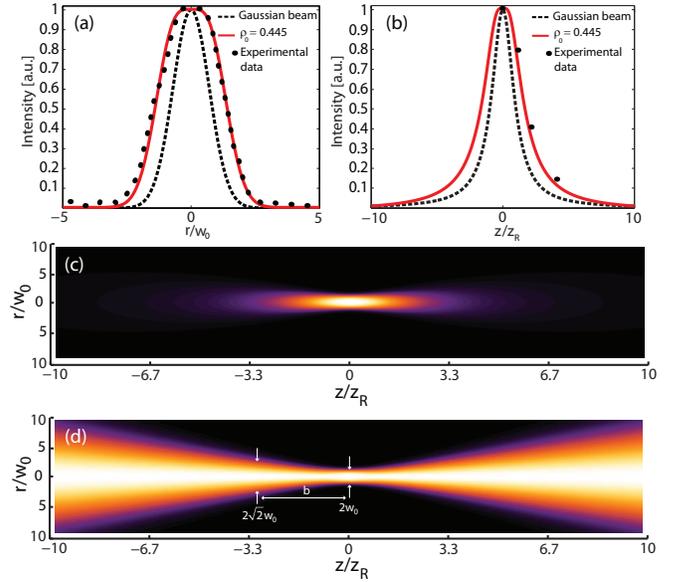}
\caption{Cross-section of the intensity distribution along (a) the radial direction and (b) the beam propagation direction at the beam center for the original Gaussian beam (black dashed line) and the super-Gaussian conical refraction (SGCR) beam obtained by adjusting $\rho_0=0.445$ (red solid line). Black dots are the corresponding experimental measurements. (c) Intensity distribution in the $z$-$r$ plane showing the free space evolution of the SGCR beam. (d) Intensity distribution of the SGCR beam normalized to the beam area to help visualizing the value of its confocal parameter \textit{b}.} 
\label{fig3}
\end{figure}

We have undertaken further investigations under the condition $\rho_0 \leq 1$ (see Figs.~\ref{fig2}(e)-(h)). We have found that at the value $\rho_0 = 0.445$ the transverse intensity profile of the transformed beam becomes flat at its top and it decays smoothly at its edges, i.e. it is a super-Gaussian conically refracted (SGCR) beam. 
To deduce such value of $\rho_0$, we solved Eqs.~(\ref{Eqs_Bc_general}) and (\ref{Eqs_Bs_general}) numerically and we looked for which value of $\rho_0$ it is found a maximum number of points with a slope equal to zero at the transverse cross-section at the focal plane.
Fig.~\ref{fig3}(a) and Fig.~\ref{fig3}(b) show, respectively, the cross-section of the transverse intensity profile of the input Gaussian beam and the output CR beam for $\rho_0=0.445$ along both the radial and the longitudinal directions. Fig.~\ref{fig3}(c) plots the 2D intensity distribution of the SGCR beam in the $z$-$r$ plane, while Fig.~\ref{fig3}(d) plots the corresponding intensity distribution normalized to the beam area. 
The plateau at the top part has been measured to be a $30\%$ of the FWHM, which is compatible with a super-Gaussian beam of first order \cite{bagini1996}. Also, the depth of field of this super-Gaussian beam is larger than that one of a Gaussian beam, as shown in Fig.~\ref{fig3}(b). In Gaussian beams, the depth of field or confocal parameter \textit{b} is twice the distance of the transverse plane at which the beam waist radius is $w(z_R) = \sqrt{2} w_0$, i.e. $b = 2 z_R$. 
To obtain the confocal parameter for the SGCR beam, we solved numerically Eqs.~(\ref{Eqs_Bc_general}) and (\ref{Eqs_Bs_general}) using $\rho_0 = 0.445$ and we looked for the axial distance $Z$ from the focal plane to which the area occupied by the SGCR beam was doubled with respect to the focal plane, i.e. we looked for Z accomplishing that $w(Z) = \sqrt{2} w(Z=0)$. The waist radius of beam, $w(Z)$, was considered at $e^{-2}$ of the maximum intensity at each plane. For the SGCR, we have found that the depth of field is $b_{\rm{SGCR}} = 6.1 z_{R}$, as depicted in Fig.~\ref{fig3}(d).
Therefore, the SGCR beam reported here has a confocal parameter three times larger than that one of the fundamental Gaussian beam. 

\begin{figure}[]
\centering
\includegraphics[width=\columnwidth]{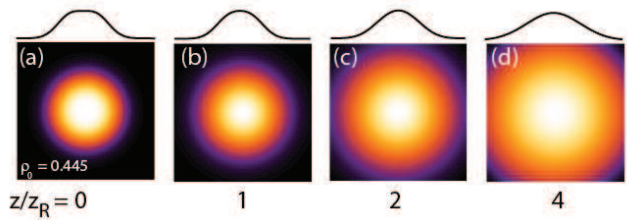}
\caption{Transverse patterns along the beam propagation direction of the SGCR beam, i.e. for $\rho_0 = 0.445$, calculated theoretically, see Eqs.~(\ref{eqI})-(\ref{Eqs_Bs_general}), for an input beam of transverse Gaussian profile. Insets are the cross-section of the intensity distribution along the horizontal direction.}
\label{fig4}
\end{figure}

Fig.~\ref{fig4} presents the theoretical transverse patterns of the SGCR beam at different planes along its propagation for a Gaussian input beam. Top insets demonstrate that the flat-top profile of the SGCR beams is only obtained at the focal plane while when the imaging plane is moved to other planes the plateau disappears and the profile tends to be Gaussian-like. 
With respect to the state of polarization (SOP) of the SGCR beam, we have found that, at the focal plane and at the beam's center, the SOP is circular, as the input beam. From this point to the ends of the pattern, the SOP evolves into the well known CR SOP. Moving from the focal plane to other planes beyond results in a transformation of the SOP into circular at any point of the transverse pattern.

\begin{figure}[]
\centering
\includegraphics[width=\columnwidth]{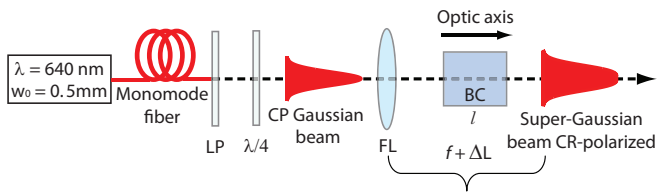}
\caption{Experimental set-up. A diode laser coupled to a monomode fiber generates a Gaussian beam at $640\,\rm{nm}$ whose polarization is fixed to be circular by using the linear polarizer (LP) and the $\lambda/4$ waveplate. Then the beam is focused by means of a focusing lens (FL) along one of the optic axes of a $\rm{KGd(WO_4)_2}$ biaxial crystal that transforms it into a super-Gaussian conical refraction beam. Experimental parameters $f=200\,\rm{mm}$, $l=2.4\,\rm{mm}$ ($R_{0}=40.8\,\rm{\mu m}$).}
\label{fig5}
\end{figure}

\begin{figure}[]
\centering
\includegraphics[width=\columnwidth]{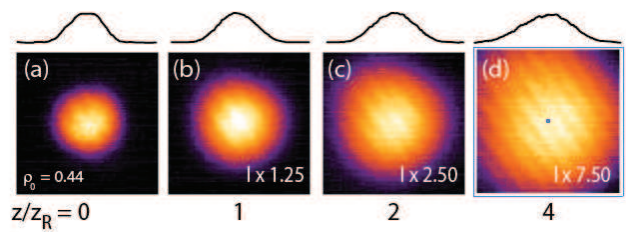}
\caption{Spatial evolution of the experimental transverse intensity patterns along the beam propagation direction for $\rho_0^{exp}=0.44$. Top insets represent the cross-section of the intensity distribution along the horizontal axis at the beam center.}
\label{fig6}
\end{figure}

Experiments on the SGCR beam only require an input beam, a BC and an imaging system (CCD camera), see Fig.~\ref{fig5}. As input beam, we take a collimated linearly polarized Gaussian beam with $w_0=0.55\,\rm{mm}$ beam waist obtained from a $640\,\rm{nm}$ diode laser coupled to a monomode fiber. The linear polarizer (LP) and the $\lambda/4$ waveplate are used to control the polarization of the input beam and fix it to be circular. To modify the input beam's waist and vary the $\rho_0$ parameter we use a plano-convex focusing lens with focal length of $200\,\rm{mm}$. As BC we use acommercially available (CROptics) $\rm{KGd(WO_4)_2}$ crystal being $l = 2.4\,\rm{mm}$, $\alpha = 17.6$~mrad and, therefore, $R_0 = 40.8\,\rm{\mu m}$. The polished entrance surfaces of the BC (cross-section $6 \times 4~\rm{mm^2}$) have parallelism with less than 10~arc sec of misalignment, and they are perpendicular to one of the two optic crystal axes within $1.5\,\rm{mrad}$ misalignment angle. To have a fine alignment of the BC, we mount it over a micro-positioner that allows modifying independently the $\theta$ and $\phi$ angles (when considering spherical coordinates) and we image the transverse CR pattern at the focal plane with a CCD camera, while slightly modifying the $\theta$ and $\phi$ angles until a cylindrically symmetric pattern is obtained. This means that the input beam passes exactly along one of the optic axis of the BC. 
The focal plane is shifted longitudinally from the source plane of the input beam by the quantity $\Delta L= l (1-1/n_{\rm{BC}})$ introduced by the BC, where $n_{\rm{BC}}$ is the mean value of the refractive indices of the BC \cite{2008_Kalkandjiev_SPIE_6994}. The beam waist of our focused Gaussian beam is measured to be $w_0 = 93\,\rm{\mu m}$
, which leads to a control parameter of $\rho_0 = 0.44$. Fig.~\ref{fig6} shows the experimental transverse patterns obtained at different propagation distances for the SGCR beams. Top insets represent the cross-section at the beam center along the horizontal direction. They agree well with the corresponding numerical simulations presented in Fig.~\ref{fig4}. For this arrangement, we have also measured the confocal parameter obtaining that $b_{\rm{SGCR}}^{exp} = (5.8 \pm 0.2) z_{R}$. 

In summary, we have demonstrated that the CR phenomenon can be used as a tool for generating beams with super-Gaussian transverse profiles, this means flat-top beams with a smooth decay at their edges, from input beams with transverse Gaussian profile. We have reported that the generation of the SGCR beams is governed by the adjustment of the control parameter $\rho_0 = R_0/w_0$ to a value of $\rho_0 = 0.445$. We have shown that the SGCR beams have only the flat-top profile at the focal plane of the system, while far from the focal plane the transverse cross-section becomes Gaussian-like with the propagation distance. Additionally, the SGCR beams have been shown to have a confocal parameter of $6.1 z_R$, i.e. more than three times that one for Gaussian beams. 
The presented method has the advantage that the CR phenomenon preserves the full power of the input beam 
as long as the facets of the crystal have dielectric coatings to avoid reflections and the doping elements that can be found in some particular crystals do not absorb input light power at the used frequency.
Additionally, this technique can be used in all the spectral range to which the biaxial crystal is transparent, at variance with other techniques such as computer generated holograms with spatial light modulators or diffractive optical elements. Also, biaxial crystals can be used with high power beams, being this essential in applications such as laser machining, optical trapping of neutral atoms and Bose-Einstein condensates and even in high intensity x-rays experiments.  
One possible drawback of the presented method is the state of polarization of the SGCR beam, which is non-uniform and contains circular, elliptical and linear SOPs. However, the SOP of the beam is only relevant in optical processing with tightly focused beams, while for the rest of the possible applications the most relevant aspect is the shape of the beam.
Finally, note that the current method could be extended to other input beams such as elliptical beams, which have been studied in detail in CR \cite{ebs2013}, to generate elliptical beams with transverse flat-top cross-section. 

The authors gratefully acknowledge financial support through Spanish MICINN contract FIS2011-23719, and the Catalan Government contract SGR2009-00347. A. Turpin acknowledges financial support through grant AP2010-2310 from the MICINN.

\end{document}